\documentclass[pdflatex,sn-mathphys-num]{sn-jnl}  

\pdfoutput=1

\usepackage{amsmath}
\usepackage{booktabs}
\usepackage{graphicx}
\usepackage{lineno}
\usepackage[version=4]{mhchem}
\usepackage{orcidlink}
\usepackage{physics2}
\usepackage{siunitx}
\usepackage{xcolor}
\usepackage{enumitem}

\DeclareSIUnit\mK{mK}
\DeclareSIUnit\sccm{SCCM}
\DeclareSIUnit\rpm{RPM}
\DeclareSIUnit\fluxquantum{\ensuremath{\Phi_0}}

\usephysicsmodule{braket}

\begin{document}
    \title{Performance Characterization of a Multi-Module Quantum Processor with Static Inter-Chip Couplers}

    \author*[1,2]{\fnm{Graham J.} \sur{Norris}\,\orcidlink{0000-0002-1016-9956}}
    \email{graham.norris@phys.ethz.ch}
    \equalcont{These authors contributed equally}

    \author*[1,2,3]{\fnm{Kieran} \sur{Dalton}\,\orcidlink{0009-0006-2416-356X}}
    \email{kieran.dalton@phys.ethz.ch}
    \equalcont{These authors contributed equally}

    \author[1,2]{\fnm{Dante} \sur{Colao~Zanuz}\,\orcidlink{0009-0001-6623-9132}}

    \author[1]{\fnm{Alexander} \sur{Rommens}\,\orcidlink{0000-0001-6142-1448}}

    \author[1,2,3]{\fnm{Alexander} \sur{Flasby}}

    \author[1,2,3]{\fnm{Mohsen} \sur{Bahrami~Panah}}

    \author[1,2]{\fnm{François} \sur{Swiadek}\,\orcidlink{0009-0000-2823-8246}}

    \author[1,2]{\fnm{Colin} \sur{Scarato}\,\orcidlink{0000-0003-3211-6334}}

    \author[1,2]{\fnm{Christoph} \sur{Hellings}\,\orcidlink{0000-0002-9929-9684}}
    
    \author[1,2,3]{\fnm{Jean-Claude} \sur{Besse}\,\orcidlink{0000-0002-1490-0072}}

    \author*[1,2,3]{\fnm{Andreas} \sur{Wallraff}\,\orcidlink{0000-0002-3476-4485}}

    \affil[1]{\orgdiv{Department of Physics}, \orgname{ETH Zürich}, \orgaddress{\postcode{CH-8093} \city{Zürich}, \country{Switzerland}}}

    \affil[2]{\orgdiv{Quantum Center}, \orgname{ETH Zürich}, \orgaddress{\postcode{CH-8093} \city{Zürich}, \country{Switzerland}}}

    \affil[3]{\orgdiv{ETH Zürich---PSI Quantum Computing Hub}, \orgname{Paul Scherrer Institut}, \orgaddress{\postcode{CH-5232} \city{Villigen}, \country{Switzerland}}}

    \abstract{
        Three-dimensional integration technologies such as flip-chip bonding are a key prerequisite to realize large-scale superconducting quantum processors.
        Modular architectures, in which circuit elements are spread over multiple chips, can further improve scalability and performance by enabling the integration of elements with different substrates or fabrication processes, by increasing the fabrication yield of completed devices, and by physically separating the qubits onto distinct modules to avoid correlated errors mediated by a common substrate.
        We present a design for a multi-chip module comprising one carrier chip and four qubit modules.
        Measuring two of the qubits, we analyze the readout performance, finding a mean three-level state-assignment error of \num{9e-3} in \qty{200}{\ns}.
        We calibrate single-qubit gates and measure a mean simultaneous randomized benchmarking error of \num{6e-4}, consistent with the coherence times of the qubits.
        Using a wiring-efficient static inter-module coupler featuring galvanic inter-chip transitions, we demonstrate a controlled-Z two-qubit gate in \qty{100}{\ns} with an error of \num{7e-3} extracted from interleaved randomized benchmarking.
        Three-dimensional integration, as presented here, will continue to contribute to improving the performance of gates and readout as well as increasing the qubit count in future superconducting quantum processors.
	}

    \maketitle

    \section{Introduction}

        Quantum computers may enable integer factorization for cryptanalysis \cite{montanaro2016quantum} and simulations of many-body condensed-matter systems for chemistry \cite{bauer2020quantum} and physics \cite{dalzell2023quantum} beyond those attainable with conventional computers.
        Given the error rates on current devices \cite{acharya2024quantum, dalton2024quantifying}, quantum error correction will be necessary to reach these goals \cite{knill1997theory}.
        Quantum error correction reduces algorithmic errors at the expense of requiring additional physical qubits.
        Thus, low-error qubit architectures which can scale to thousands of qubits \cite{vonburg2021quantum,gidney2021how} are needed.

        Superconducting qubits are a leading platform for building large-scale processors due to their lithographic micro-/nano-fabrication, high coupling rates resulting in short gate durations, and demonstrated low-error operations \cite{kjaergaard2020superconducting,blais2021circuit}.
        Three-dimensional (3D) integration techniques for superconducting circuits, such as flip-chip bonding \cite{rosenberg2017three, foxen2020demonstrating, gold2021entanglement} and out-of-plane wiring \cite{bejanin2016three,bronn2018high, rahamim2017double, nakamura2023superconducting}, can increase the number of qubits which can be included in a single quantum processor \cite{rosenberg2020solid}.

        While conventional flip-chip bonded devices comprise two chips bonded together, \emph{modular} devices separate the circuit components across multiple chips (forming a multi-chip module), providing a number of advantages.
        Firstly, modular designs allow minimal, optimized fabrication procedures to be used for each chip depending on its functionality, possibly improving coherence times \cite{putterman2024hardware}.
        In addition, this may enable the integration of other qubit modalities such as semiconducting quantum dots \cite{holman2021three}, nano-mechanical resonators \cite{wollack2022quantum}, acoustic resonators \cite{chu2018creation}, optical-to-microwave transducers \cite{brubaker2022optomechanical}, or single-flux quantum \cite{liu2023single} devices for \emph{in situ} control.
        Further, modular devices with fewer qubits per module are expected to improve fabrication yield over many-qubit planar or two-chip 3D-integrated devices.
        As chip sizes and qubit counts grow, the chance of a photolithography defect or qubit frequency targeting error also increases.
        One such defect can render the entire chip inoperative, resulting in a low fraction of functional devices, or \emph{yield}.
        If the device is split up over several smaller modules, then the presence of a defect only requires that one module is replaced rather than the entire device, increasing the yield \cite{smith2022scaling}.
        Separating the qubits onto distinct substrates should also eliminate correlated errors caused by charge carriers or phonons propagating through the substrate \cite{martinis2021saving,wilen2021correlated,mcewen2022resolving}.
        Additionally, modular construction \cite{bravyi2022future} may enable scaling beyond the current limits of individual samples (\textit{circa} $10^3 $ qubits \cite{castelvecchi2023ibm}) or cryostats (\textit{circa} $10^5$ qubits), through inter-sample \cite{kurpiers2018deterministic,dickel2018chip,zhong2021deterministic,niu2023low} and inter-cryostat \cite{magnard2020microwave} links.

        Universal quantum computation requires high-fidelity readout and single- and two-qubit gate operations.
        In particular, for superconducting circuits, two-qubit gates currently limit algorithm performance \cite{krinner2020benchmarking,acharya2024quantum}.
        3D-integrated two-qubit gates between transmon qubits have been studied on flip-chip devices comprising two chips bonded together (two-chip 3D-integrated devices) using static inductive \cite{connor2021superconducting} and tunable \cite{foxen2020demonstrating, kosen2022building} couplers.
        Some gate error information is also available from commercial 3D-integrated devices \cite{arute2019quantum,jurcevic2021demonstration,acharya2024quantum,abdurakhimov2024technology}, but these works provide little detail about the device architectures or gate implementations.

        For the reasons discussed above, modular devices are a promising approach to scaling 3D-integrated quantum processors.
        However, previous two-qubit gates demonstrated in multi-chip modules suffered from high susceptibility to inter-chip separation \cite{gold2021entanglement} or required SQUID-based coupling elements with additional control lines \cite{field2024modular, wu2024modular}.
        In this work, we have developed and characterized an architecture based on frequency tunable transmon qubits and wiring-efficient inter-module static couplers featuring galvanic indium bump bonds to route signals between chips.
        We describe the architectural and device parameter choices and present the complete fabrication procedure in Sect.~\ref{sec:main:design}.
        In Sect.~\ref{sec:main:qubit-frequency-coherence}, we characterize the frequencies and coherence times of our qubits.
        We then examine the performance of our readout circuitry featuring dedicated Purcell filters \cite{heinsoo2018rapid} in Sect.~\ref{sec:main:readout}.
        We analyze the qubit-qubit coupling rate and the performance of single- and two-qubit gates using randomized benchmarking in Sect.~\ref{sec:main:gates} before discussing our findings in Sect.~\ref{sec:main:discussion}.
    
    \section{Sample design}
        \label{sec:main:design}

        A conventional (planar) device consists of a single substrate with a layer of patterned superconductor on its top face.
        A flip-chip 3D-integrated device adds an extra substrate and superconducting layer which enables the distribution of circuit elements, avoiding signal routing bottlenecks and reducing crosstalk \cite{karamlou2024probing,kosen2024signal}.
        However, this flexibility comes at a cost: capacitive or inductive inter-chip coupling rates depend on the inter-chip separation.
        Previous work on flip-chip devices has found significant separation variations \cite{foxen2018qubit,kosen2022building,norris2024improved} which can negatively impact two-qubit gate performance \cite{gold2021entanglement}.
        To avoid these coupling rate variations, we employ three strategies.
        First, we use galvanic inter-chip coupling where critical.
        If designed well, a galvanic connection should provide separation-independent coupling \cite{field2024modular}, although they may introduce loss depending on implementation \cite{rosenberg2017three}.
        Second, we prefer intra- rather than inter-chip capacitive coupling.
        Due to the large disparity between the dielectric constant of the substrate (here silicon, $\epsilon_\mathrm{r} = 11.45$) and vacuum, the intra-chip capacitance is typically dominated by the field in the substrate rather than in the vacuum gap between the chips.
        Third, we use mechanical spacers \cite{norris2024improved,niedzielski2019silicon,li2021vacuum} to suppress inter-chip separation variations, addressing the root cause of the coupling rate variations.
        
\begin{figure*}
            \centering
            \includegraphics[width=372pt]{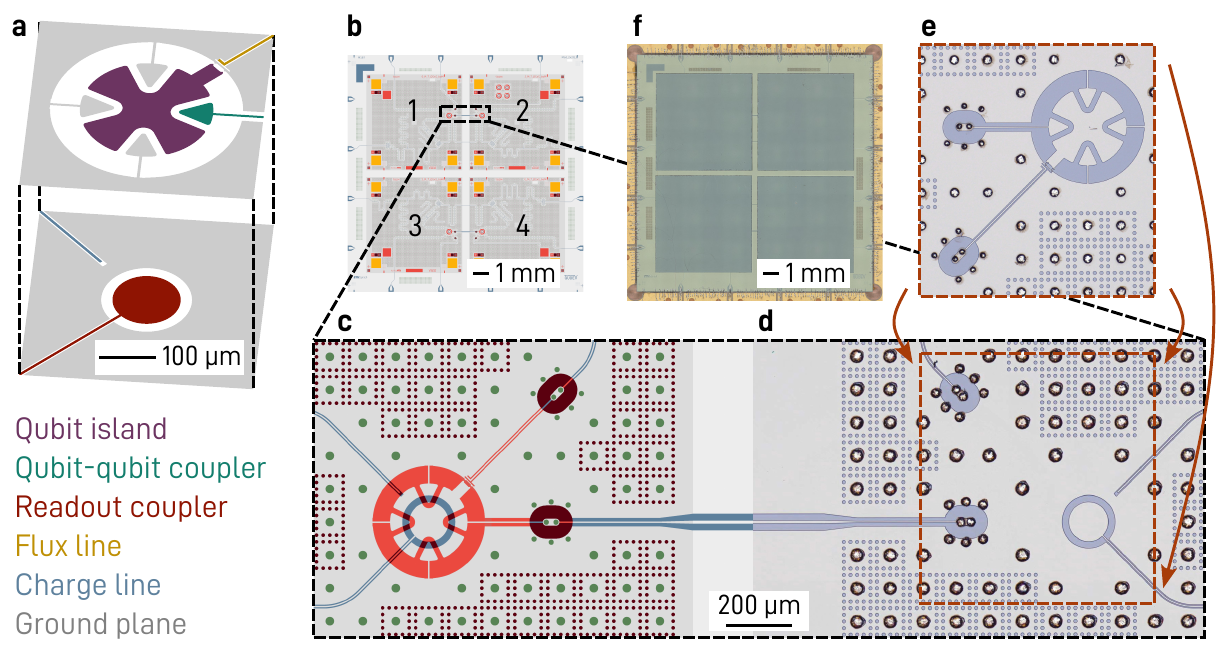}
            \caption{
                Device design and implementation.
                (a) Schematic of the qubit design showing the island (purple), qubit-qubit coupler (green), readout coupling pad (red), flux line (yellow), charge line (blue), and ground plane (gray).
                Metal on each chip is indicated by the presence of shaded regions.
                The vertical separation between the chips of \qty{10}{\um} is not to scale.
                (b) Render of the device design.
                Niobium is shown as light gray on the carrier and darker gray on the modules.
                Areas of exposed silicon on the carrier are rendered in blue while equivalent areas on the modules are rendered in light red.
                Indium bumps are shown in green and SU-8 spacers are colored in yellow.
                Subtractive color is used, so dark red indicates overlapping areas of exposed silicon on both chips.
                (c) Detail of the device design render showing a qubit and half the qubit-qubit coupler.
                This qubit-qubit coupler includes a galvanic inter-chip coupler and high-impedance coplanar waveguides to increase the coupling rate.
                (d) Optical micrograph of the carrier region symmetric to the part of the device rendered in (c).
                (e) Optical micrograph of the matching module region with the qubit island.
                (f) Optical micrograph of the completed multi-chip module.
            }
            \label{fig:main:introduction}
        \end{figure*}

        We place the islands of our transmon qubits on the top chips (modules) to separate them from the readout circuitry and charge lines on the bottom chip (carrier), see Fig.~\ref{fig:main:introduction}a.
        We target an anharmonicity of $\alpha/2\pi=\qty{-165}{\MHz}$ to balance single qubit gate duration \cite{lazar2023calibration} against charge dispersion \cite{koch2007charge}.
        On one side, the island (purple) features a metallic extension close to the ground plane where both junctions of an asymmetric superconducting quantum interference device (SQUID) loop are grounded (not depicted in the diagram).

        We control the flux threading the SQUID loop by introducing a flux control line (yellow) which is shorted to ground adjacent to one edge of the loop.
        To drive the qubit transitions, we use weak capacitive coupling to a charge line (blue) on the wiring chip and we employ a large pad (red) directly beneath the qubit island to couple it to the readout circuitry.

        All elements are designed to reduce the sensitivity of intra-chip coupling rates to lateral misalignment of the module relative to the carrier by several \si{\um}.
        The flux line is located on the module, so the mutual inductance of the line to the SQUID loop should be independent of any lateral misalignment between the chips.
        We note that placing the grounded end of the flux line on the wiring chip underneath the SQUID loop may increase the mutual inductance between the line and the loop  \cite{kosen2022building}.
        Due to the weakness of inter-chip capacitance across the vacuum gap, the coupling pad for the readout resonator is large and thus should be unaffected by small lateral misalignments.

        We implement this qubit design in a multi-chip module comprising one \qty{14.3}{\mm} by \qty{14.3}{\mm} carrier chip and four \qty{5.8}{\mm} by \qty{5.8}{\mm} modules with a single qubit each, see Fig.~\ref{fig:main:introduction}b.
        While the qubits would fit on smaller modules, we use this size to simplify mechanical handling.
        We laterally separate the modules by \qty{400}{\um} to balance the attainable qubit-qubit coupling rates (discussed below) and the mechanical precision required to assemble the device.

        All control signals are routed to the perimeter of the carrier chip using coplanar waveguides.
        Since the flux lines terminate on the modules and we apply a static current to adjust the qubit frequency, we need a galvanic inter-chip connection.
        As target parameters, this connection should have a critical current sufficiently high to apply several flux quanta to the SQUID loop (order \qty{1}{\mA} \cite{krinner2019engineering}) and must be impedance matched to the waveguide to avoid unwanted reflections (ideally \qty{-20}{\dB} or below).
        We implement these connections using the design shown in Fig.~\ref{fig:main:introduction}c.
        The galvanic coupler features two \qty{15}{\um} diameter indium bumps along the transmission line for redundancy and increased spacing to the ground plane to reduce the capacitive loading from the pads.
        We depict the physical version of the coupler in Fig.~\ref{fig:main:introduction}d,e.

        Two-qubit gates are typically implemented using either tunable or static couplers between neighboring qubits.
        Tunable couplers \cite{chen2014qubit,collodo2020implementation} can enable fast and high-fidelity two-qubit gates \cite{foxen2020demonstrating,sung2021realization,marxer2023long}.
        However, they require more control lines than two-qubit gate schemes that rely only on control of the qubit frequencies\footnote{Since tunable couplers require at least one flux-bias line per qubit-qubit coupler.
        This means that, for a typical two-dimensional $n \times n$ qubit square lattice, tunable couplers require $2n(n-1)$ total flux lines (using fixed-frequency qubits) while frequency-tunable qubits and static couplers require only $n^2$ flux lines.}.
        Thus, for large-scale devices, we are motivated to use control-wiring-efficient static couplers.
        
        We implement capacitive qubit-qubit couplers using sub-wavelength transmission-line segments.
        To route the coplanar waveguide from one module to the next, we employ the same galvanic inter-chip couplers discussed above and a segment of coplanar waveguide on the carrier chip, see Fig.~\ref{fig:main:introduction}c-e.        
        We target a qubit-qubit coupling rate $2J_\text{qq}/2\pi$ of approximately \qty{14}{\MHz} to balance two-qubit gate duration against residual ZZ errors when not performing two-qubit gates \cite{krinner2020benchmarking}.
        To reach this rate at a qubit-qubit separation of \qty{2}{\mm}, we use a high-impedance coplanar waveguide (approximately \qty{70}{\ohm} instead of \qty{50}{\ohm}; estimated using conformal mapping) which reduces the capacitance to ground of the coupler, increasing the coupling rate \cite{blais2021circuit}.
        
        We fabricate our multi-chip module on a high-resistivity silicon substrate with niobium base metallization, SU-8 spacers on the carrier chip, Al/AlO$_x$/Al Josephson junctions on the modules, and indium bumps on both chips, see Fig.~\ref{fig:main:introduction}d,e.
        Further details about the fabrication procedure can be found in Appendix~\ref{sec:appendix:fabrication} and in Refs.~\cite{norris2024improved,zanuz2024mitigating}.

        We wire-bond the sample to a microwave printed-circuit board (Fig.~\ref{fig:main:introduction}f), enclose it in a metal sample package, and install the package in a dilution refrigerator with a base temperature of approximately \qty{15}{\mK}.
        The sample mount includes three layers of magnetic shields: one of high-purity aluminum and two of a high magnetic permeability nickel-iron alloy (from inside to outside).
        We discuss the measurement apparatus further in Appendix~\ref{sec:appendix:measurementapparatus}.

    \section{Qubit frequencies and coherence times}
        \label{sec:main:qubit-frequency-coherence}

        \begin{figure}
            \centering
            \includegraphics[width=125mm]{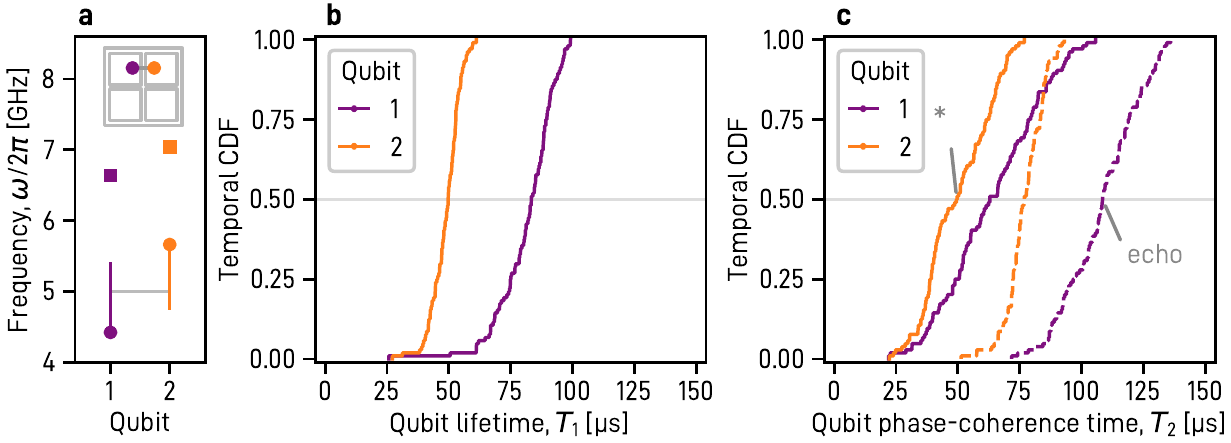}
            \caption{
                Qubit frequencies and coherence times.
                (a) Device A frequency configuration.
                Qubit idling frequencies are shown as solid circles, tunability ranges are indicated by vertical lines, and readout resonator frequencies are indicated by squares.
                The horizontal line indicates the interaction frequency at which we perform the two-qubit gate.
                Qubit positions on two of the four modules of the the device are indicated in the small diagram inset at the top of the plot; not to scale.
                (b) Measured temporal cumulative distribution function (CDF) of the qubit lifetime, $T_1$.
                (c) Measured temporal CDF of qubit Ramsey ($T_{2}^{*}$; solid lines) and Hahn ($T_{2}^{\mathrm{echo}}$; dashed lines) phase-coherence times.
            }
            \label{fig:main:qubit-frequencies}
    	\end{figure}

        We fabricated two sample designs which feature different readout circuitry configurations based on alternative choices of empirical design parameters.
        In the remainder of this work, we will primarily consider a sample of one design (device A).
        However, as we will discuss in Section~\ref{sec:main:readout}, device A has poorly targeted readout parameters, so we will characterize a sample of the other design there (device B).

        We analyze qubits 1 and 2 from device A.
        We tune qubit 1 to its minimum frequency of \qty{4.421}{\GHz} and qubit 2 to its maximum of \qty{5.662}{\GHz} where they are first-order insensitive to flux noise, see the purple (qubit 1) and orange (qubit 2) filled circles in  Fig.~\ref{fig:main:qubit-frequencies}a.
        Here, we refer to the first three states of our transmon qubits as $\ket{\mathrm{g}}$, $\ket{\mathrm{e}}$, and $\ket{\mathrm{f}}$.
        Further qubit parameters of device A are listed in Appendix~\ref{sec:appendix:hamiltonian-parameters}.

        Superconducting qubit coherence times are known to fluctuate over time \cite{burnett2019decoherence}, so individual measurements are insufficient.
        We measure the coherence times of qubits 1 and 2 104 times over 16 hours, finding median $T_1$ times of \qty{83}{\us} and \qty{50}{\us} respectively (Fig.~\ref{fig:main:qubit-frequencies}b), median $T_{2}^{*}$  times of \qty{63}{\us} and \qty{50}{\us}, and $T_{2}^{\mathrm{echo}}$ times of \qty{109}{\us} and \qty{77}{\us} (Fig.~\ref{fig:main:qubit-frequencies}c).
        For the other two qubits on device A, 3 and 4, we measure $T_1$ values of \qty{37}{\us} and \qty{53}{\us} respectively; $T_{2}^{*}$ values of \qty{47}{\us} and \qty{18}{\us}; and $T_{2}^\mathrm{echo}$ values of \qty{62}{\us} and \qty{71}{\us}, all from individual measurements (rather than time-averaged ones; further details in Appendix~\ref{sec:appendix:hamiltonian-parameters}).

        This mean $T_1$ falls within the range of commercial two-chip 3D-integrated devices \cite{jurcevic2021demonstration,acharya2024quantum,abdurakhimov2024technology}, and is comparable to other 3D-integrated qubits with known fabrication procedures---slightly below fixed-frequency qubits \cite{kosen2022building,kosen2024signal} and matching or exceeding other frequency-tunable qubits \cite{karamlou2024probing,rosen2024flat} including on other modular devices \cite{gold2021entanglement,field2024modular}.
        The high $T_1$ time of qubit 1 indicates that the presence of SU-8 spacers on the wiring chip and indium bumps on the qubit chips does not limit coherence times below this value.

    \section{Qubit readout}
        \label{sec:main:readout}

        Readout fidelity is crucial to quantum algorithms and readout duration is particularly important for error-correction \cite{krinner2022realizing,acharya2024quantum} and experiments involving classical feedback \cite{steffen2013deterministic, song2024realization}.
        As flip-chip bonding can introduce frequency targeting issues in the readout circuitry \cite{norris2024improved}, particularly in modular devices where each module might have a different inter-chip separation, we want to verify that the architecture described in Section~\ref{sec:main:design} achieves fast, high-fidelity readout.

        We use a frequency-multiplexed scheme with individual Purcell filters attached to each qubit readout resonator \cite{heinsoo2018rapid}.
        These resonators, with frequencies from \qtyrange{6.6}{7.1}{\GHz}, are roughly \qty{1.3}{\GHz} above the qubit maximum frequency to avoid Purcell decay \cite{sete2014purcell} and to place them within the \qtyrange{6}{7.5}{\GHz} frequency band of the traveling-wave parametric amplifiers (TWPAs) \cite{macklin2015near} used in these experiments.

        For the device considered so far (device A), the design parameters resulted in an approximately \qty{57}{\MHz} detuning between the Purcell-filter and dressed readout-resonator modes, leading to small readout-resonator linewidths \cite{heinsoo2018rapid} and reduced readout speed \cite{blais2021circuit}.
        Thus, for readout performance, we analyze the nominally identical qubits 1 and 2 from a device (device B) of the other design with better alignment of the resonator and filter.
        This device was co-fabricated on the same wafers as device A.

\begin{figure}
            \centering
            \includegraphics[width=125mm]{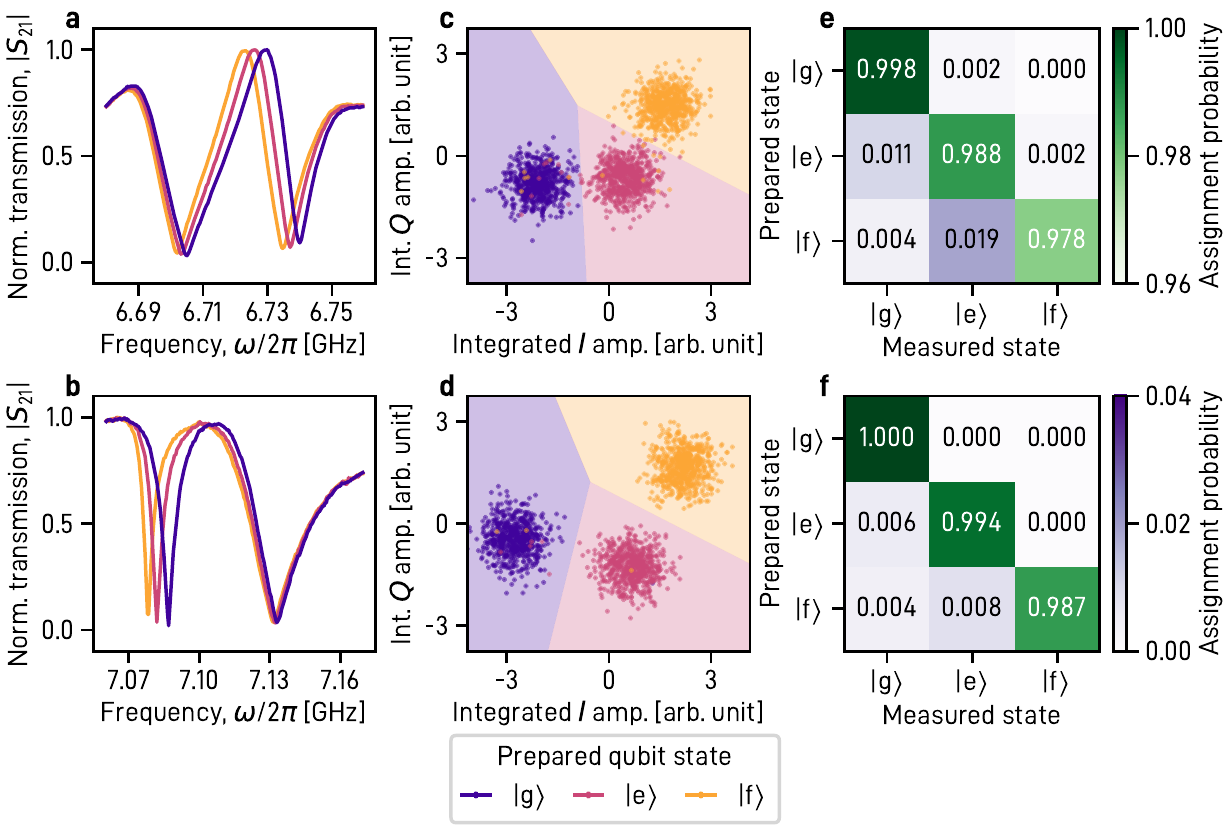}
            \caption{
                Characterization of readout performance.
                Normalized absolute transmission through the feedline of the readout-resonator Purcell-filter pair of (a) qubit 1 and (b) qubit 2 of device B for qubit states $\ket{\mathrm{g}}$, $\ket{\mathrm{e}}$, and $\ket{\mathrm{f}}$.
                (c, qubit 1; d, qubit 2) Histograms of the integrated $IQ$ voltages from single-shot measurements of the resonator response when the qubit is prepared in $\ket{\mathrm{g}}$, $\ket{\mathrm{e}}$, and $\ket{\mathrm{f}}$.
                One in every 15 measurements per state is shown.
                The shaded regions indicate the most likely state per integrated $IQ$ value based on a Gaussian mixture model.
                (e, qubit 1; f, qubit 2) State-assignment probability matrices after ground-state pre-selection and classifying the integrated response.
            }
            \label{fig:main:readout-performance}
    	\end{figure}

        We measure the transmission through the feedline and find the two readout-resonator and Purcell-filter pairs shown in Fig~\ref{fig:main:readout-performance}a,b.
        Each plot features a Purcell-filter-like mode (broader linewidth, smaller dispersive shift) and a readout-resonator-like mode (narrower linewidth, larger dispersive shift) with a resonator-filter detuning of approximately \qty{20}{\MHz}.
        We find that the readout resonator is above the Purcell filter for qubit 1 and below for qubit 2, contrary to the design parameters, where both should be on resonance.
        This small detuning (approximately half the Purcell-filter linewidth) results in a broad readout-resonator linewidth of $\kappa_\mathrm{r}/2\pi = \qty{7.5}{\MHz}$ (\qty{11.2}{\MHz}) for qubit 1 (2), as desired for fast readout.
        We have also measured a dispersive shift, $2\chi/2\pi = (\omega_\mathrm{r,e} - \omega_\mathrm{r,g})/2\pi$, the difference between the $\ket{\mathrm{e}}$ and $\ket{\mathrm{g}}$ state responses, of \qty{-4.4}{\MHz} (\qty{-6.2}{\MHz}) for the qubit 1 (2) readout-resonator-like mode.
        These values are reasonably close to the $\kappa \approx 2 \lvert\chi\rvert$ steady-state ideal derived for a simpler system \cite{gambetta2008quantum}.
        For these measurements, we statically tune qubit 1 from its frequency minimum to \qty{5.0}{\GHz} to increase the dispersive shift.
        The dependence of readout performance on the detuning between the qubit and the readout resonator is analyzed in detail in Ref.~\cite{swiadek2024enhancing}.

        We apply a readout tone at the frequency which maximizes the sum of the differences in absolute transmission between the ($\ket{\mathrm{g}}$, $\ket{\mathrm{e}}$) and ($\ket{\mathrm{e}}$,  $\ket{\mathrm{f}}$) state responses.
        We integrate the response using nearly-optimal linear weights \cite{gambetta2007protocols} adapted to three levels \cite{krinner2022realizing}.
        For qubit 1 (2), we use an integration time of \qty{192}{\ns} (\qty{160}{\ns}), balancing increased signal against $T_1$ decay of the qubit \cite{blais2021circuit}.
        
        Repeating the readout $10^4$ times per prepared state results in the histograms shown in Fig.~\ref{fig:main:readout-performance}c,d where the qubit states are well-resolved in the plane of the two detection quadratures.
        To determine the qubit state, we threshold the integrated $IQ$ voltage based on a Gaussian mixture model as indicated by the shaded regions in the figure.
        If we add a pre-selection readout before the start of each experiment, then we can use these thresholds to reject instances where the qubit is not in $\ket{\mathrm{g}}$ prior to the state-preparation pulses, discarding \qty{0.3}{\percent} (\qty{0.2}{\percent}) of the repetitions for qubit 1 (2).
        After pre-selection, we find the state-assignment probability matrices in Fig.~\ref{fig:main:readout-performance}e,f.

        Based on these measurements, we calculate a mean state-assignment error of \num{1.2e-2} (\num{6e-3}) for qubit 1 (2), comparable to the fidelities reported in large-scale devices \cite{krinner2022realizing,acharya2024quantum}.
        Our results concur with those of Ref.~\cite{spring2024fast}, showing that high-fidelity readout of modular 3D-integrated devices is achievable.
        We provide details on the readout parameters of both samples in Appendix~\ref{sec:appendix:readout-parameters}.

    \section{Single- and two-qubit gates}
        \label{sec:main:gates}

        \begin{figure}
            \centering
            \includegraphics[width=125mm]{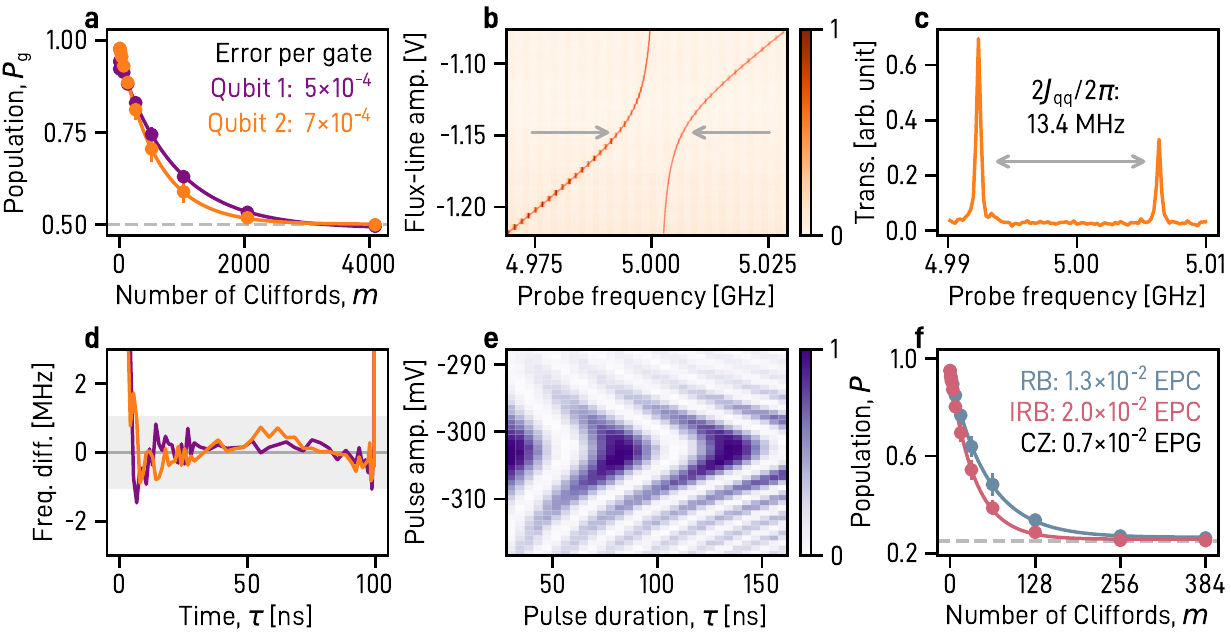}
            \caption{
                Characterization of single- and two-qubit gate performance of device A.
                (a) Randomized benchmarking (RB) of the single-qubit gates of qubits 1 and 2.
                Here and in (f), the vertical error bars at each point and uncertainties in the fit values are discussed in the text and the gray dashed line indicates the population expected for a fully mixed state.
                (b) Two-tone spectroscopy measurements of qubit 2 as it is tuned through resonance with qubit 1 in the single-excitation manifold.
                The fit is shown as a semi-transparent red line on top of the measured data.
                (c) Transmission data at the flux line voltage indicated by the arrows in panel (c) showing the extracted qubit-qubit coupling rate.
                (d) Qubit frequencies during a \qty{100}{\ns} flux pulse relative to the mean extracted using the cryoscope method \cite{rol2020time}.
                For this measurement, the pulsed qubit idles at its maximum frequency and the other qubit idles at its minimum.
                (e) Measurement of the $\ket{\mathrm{ee}}$ state swapping into $\ket{\mathrm{gf}}$ with the two states near resonance.
                The purple color indicates the qubit 1 $\ket{\mathrm{g}}$-state population as a function of the duration of the pulses on both qubits and amplitude of the qubit 2 pulse.
                (f) RB and interleaved RB (IRB) measurements of the CZ gate with extracted error per Clifford (EPC) and per gate (EPG).
                The y-axis shows the $\ket{\mathrm{gg}}$ population.
            }
            \label{fig:main:two-qubit-gate-performance}
    	\end{figure}

        Gate performance is critical for quantum algorithms and currently two-qubit gates are one of the dominant error sources for large quantum experiments \cite{krinner2022realizing,acharya2024quantum}.
        Thus, we have carefully analyzed the performance of single- and two-qubit gates on device A.
    
        We calibrate single-qubit gates with squared-cosine envelopes on qubits 1 and 2 using separate measurements for $\pi$ and $\pi/2$ rotations \cite{lazar2023calibration}.
        We characterize their performance with randomized benchmarking (RB) \cite{magesan2012characterizing} using an XZ decomposition \cite{lazar2023improving} to take advantage of virtual-Z gates \cite{mckay2017efficient}.
        We apply varying-length sequences of random Clifford operations simultaneously \cite{gambetta2012characterization} to both qubits and extract an error per gate of \num{5.0(2)e-4} for qubit 1 (purple) and \num{7.4(0.1)e-4}) for qubit 2 (orange) from the exponential decays of the ground state population in Fig.~\ref{fig:main:two-qubit-gate-performance}a.
        The vertical error bars at each point represent the standard deviation of the data and the quoted uncertainty of the error per gate represents the statistical uncertainty of the fit.
        Using the median coherence times of the qubits (Fig.~\ref{fig:main:qubit-frequencies}b,c) and the methods of Ref.~\cite{abad2022universal}, we compute coherence limits of \num{3.6e-4} and \num{5.0e-4} respectively, near our measured gate errors.

        These single-qubit gate errors are slightly worse than those published for simultaneous RB in two-chip 3D-integrated devices with fixed-frequency qubits \cite{kosen2022building}, which we attribute to our slightly lower coherence times and longer single-qubit gate duration.
	    We also note that devices with static couplers may be affected by spectator errors due to residual ZZ coupling of the qubits \cite{krinner2020benchmarking}.
        
        Next, we analyze the inter-module qubit-qubit coupling.
        Using two-tone spectroscopy \cite{blais2021circuit} of qubit 2 while sweeping the flux through its SQUID loop, we identify an avoided crossing in the $\ket{\mathrm{ge}}$-$\ket{\mathrm{eg}}$ manifold with both qubits tuned near resonance at \qty{5}{\GHz}, see Fig.~\ref{fig:main:two-qubit-gate-performance}b.
        We extract a qubit-qubit coupling rate $2J_\mathrm{qq}/2\pi$ of \qty{13.4(0.1)}{\MHz}, Fig.~\ref{fig:main:two-qubit-gate-performance}c.
        This is in reasonable agreement with a simulated qubit-qubit coupling rate of \qty{13.6}{\MHz}.
        To attain this value, we use numerical admittance-matrix calculations combining capacitance values from finite-element simulations with coplanar waveguide properties from conformal mapping.

        Two-qubit gates relying on fast flux-mediated qubit frequency control \cite{strauch2003quantum,dicarlo2009demonstration} are susceptible to distortions in the base-band flux pulses caused by reactances in the flux-bias lines \cite{rol2019fast}.
        These distortions must be measured and compensated for to realize high-fidelity gates.
        We use a pulsed qubit spectroscopy method \cite{krinner2022realizing,hellings2025calibrating} to measure how the qubit frequency reacts to an applied step-pulse in the flux-control line over time scales from \qty{100}{\ns} to \qty{100}{\us} and counteract distortions with infinite-impulse-response filters.
        We then characterize the flux-line transfer function for times shorter than \qty{100}{\ns} using the cryoscope method \cite{rol2020time} and correct the response using finite-impulse-response filters \cite{hellings2025calibrating}.
        After activating these pre-distortion filters, we find qubit frequency variations of less than roughly \qty{1}{\MHz} during a \qty{100}{\ns} flux pulse when driving the qubits \qty{200}{\MHz} below their maximum frequencies, see Fig.~\ref{fig:main:two-qubit-gate-performance}d.
        We do not find that a different number or character of filters is required to reach this level of flux-pulse frequency control compared to planar devices without the galvanic inter-chip coupler.
        Thus, given the small remaining variations, we conclude that the galvanic inter-chip coupler and 3D-integration of the flux line introduces only flux pulse distortions which are addressable with conventional approaches (if any).
        
        With the flux-pulse pre-distortion filters in place, we apply unipolar flux pulses to both qubits designed to tune the qubit 2 $\ket{\mathrm{e}}$-$\ket{\mathrm{f}}$ transition into resonance with the qubit 1 $\ket{\mathrm{g}}$-$\ket{\mathrm{e}}$ transition at an interaction frequency of \qty{5.0}{\GHz}.
        We sweep the duration of the pulse applied to both qubits and the amplitude of the pulse applied to qubit 2 and measure population swapping from $\ket{\mathrm{ee}}$ to $\ket{\mathrm{gf}}$, resulting in the characteristic response shown in Fig.~\ref{fig:main:two-qubit-gate-performance}e.
        The shape is symmetric, indicating minimal distortions over the observed time period \cite{rol2020time} and we find a population recovery time of approximately \qty{55}{\ns}, close to the minimum gate duration of $\pi/\sqrt{2} J_\mathrm{qq} = \qty{53}{\ns}$ set by our qubit-qubit coupling rate.

        To implement the CZ gate, we choose a pulse duration near the first minimum of the $P_\text{g}$ response in Fig.~\ref{fig:main:two-qubit-gate-performance}e, and calibrate the amplitude and duration of a net-zero pulse shape \cite{negirneac2021high} to minimize leakage and provide a conditional phase of $\pi$.
        We additionally measure the dynamic phase acquired by the other states during the frequency excursion and use virtual-Z gates to update their rotating frames accordingly.
        We note that the net-zero shape minimizes memory effects during repeated gate applications and provides an echo-pulse-like effect \cite{negirneac2021high} which reduces dephasing during the gate.
        The final gate has a total duration of \qty{103}{\ns} (including \qty{20}{\ns} buffers on either side).

        With this CZ gate calibrated, we analyze its performance using interleaved randomized benchmarking (IRB) \cite{magesan2012efficient}.
        We perform standard two-qubit RB and then the interleaved form, with the CZ gate inserted between each random two-qubit Clifford operation.
        As in the single-qubit case, we observe exponentially decaying ground state populations and extract an error per two-qubit Clifford of \num{1.34(3)e-2} and \num{2.02(4)e-2} for RB and IRB, respectively, resulting in an error per CZ gate of \num{7.0(5)e-3}, see Fig.~\ref{fig:main:two-qubit-gate-performance}f.
        Using similar methods, we find an $\ket{\mathrm{f}}$-level leakage per CZ gate of \num{4(1)e-4}.
        We use the methods of Ref.~\cite{abad2022universal} and estimate a coherence-limited gate error of \num{8.9e-3} using the measured $T_{2}^{\mathrm{echo}}$ times at the interaction frequency for the duration of the flux-pulse and the $T_{2}^{*}$ times at the idling frequencies for the buffers on either side.

        The low gate error further indicates that the two galvanic inter-chip couplers are not detrimental to gate performance at this level, as this fidelity is similar to the best we have observed on planar devices \cite{krinner2022realizing} using similar fabrication and calibration procedures.
        Compared to other modular 3D-integrated two-qubit gates, our gate error is lower than those demonstrated with parametric CZ gates using static capacitive \cite{gold2021entanglement} or tunable couplers \cite{field2024modular} likely due to the increased coherence times of our qubits.
	    This gate error is also comparable to those of other two-chip 3D-integrated devices \cite{kosen2022building,jurcevic2021demonstration,acharya2024quantum}.
        We note that increasing the number of qubits on a device tends to increase the gate errors due to effects like cross-talk or spectator errors \cite{krinner2020benchmarking}.

    \section{Discussion}
        \label{sec:main:discussion}

        In this work, we presented the design, the parameter choices, and the fabrication procedure of our modular 3D-integrated device featuring galvanic inter-chip connections.
        We found qubit coherence times comparable to those of planar devices produced in our laboratory using a similar process, and other 3D-integrated devices.
        This indicates that, as previously posited \cite{norris2024improved}, the use of SU-8 spacers and deposited indium on the qubit chips do not limit qubit coherence at the currently observed level.
        Similarly, we validated that individual Purcell filters can be used in a modular architecture to achieve fast, high-fidelity readout.

        Most importantly, we demonstrated that galvanic coupling is suitable for high-fidelity two-qubit gates, without the need for additional control lines.
        We argued that this architecture should demonstrate reduced sensitivity to inter-chip separation variations.
        While we only present a single coupling rate, its proximity to the design value provides some evidence that this process is well-targeted.
        Furthermore, we found that the galvanic couplers do not reduce the performance of two-qubit control.
        The high measured gate fidelity indicates that the galvanic coupler has low-loss.

        Future work can extend this architecture to confirm the advantages of 3D-integrated and modular devices via targeted study, particularly the roughly order-of-magnitude reduction in cross-talk discussed in the literature \cite{karamlou2024probing,kosen2024signal}, the reductions in correlated errors, and the expected improvement of fabrication yield.
        On the fabrication side, investigating reworkable flip-chip bonding procedures \cite{das2024reworkable} could allow us to combine known-good modules into functional devices.
        Finally, there is the challenge of developing high-density, low-loss connections between multi-chip modules within a package \cite{bravyi2022future}, between packages \cite{kurpiers2018deterministic,dickel2018chip,zhong2021deterministic,niu2023low}, and between cryostats\cite{magnard2020microwave}.

    \begin{appendices}

    \section{Fabrication}
        \label{sec:appendix:fabrication}

        The niobium and SU-8 patterning procedures used for the devices discussed in this paper are described in Ref.~\cite{norris2024improved}.
        The Josephson junction fabrication process and the indium deposition used for the devices considered here are described in Appendices ~\ref{sec:appendix:fabrication:junctions} and \ref{sec:appendix:fabrication:indium}.
    
        \subsection{Josephson junction fabrication}
        \label{sec:appendix:fabrication:junctions}

        We pattern Al/AlO$_x$/Al Josephson junctions using a bridge-free process \cite{lecocq2011junction} on groups of four modules.
        To avoid ion-milling induced damage \cite{nersisyan2019manufacturing,vandamme2023argon} beneath the junction leads, we deposit the entire junctions on the exposed silicon surface and then add bandages \cite{dunsworth2017characterization} in a separate step afterwards.
        
        To fabricate the leads of the Josephson junctions, we employ a bilayer resist stack.
        The lower layer ($\approx \qty{800}{\nm}$ thickness) is a copolymer of methyl methacrylate (MMA) and \qty{8.5}{\percent} methacrylic acid (MAA) diluted to \qty{12}{\percent} concentration in ethyl lactate (EL).
        The resist is spun at \qty{2000}{\rpm} (\qty{3}{\s} ramp, \qty{90}{\s} hold) and baked for \qty{300}{\s} at \qty{180}{\degreeCelsius} on top of a carrier wafer on a vacuum hotplate.
        The upper layer ($\approx \qty{200}{\nm}$ thickness) is CSAR62 ARP6200.09 which is spun at \qty{4000}{\rpm} (\qty{3}{\s} ramp, \qty{60}{\s} hold) and then baked for \qty{240}{\s} at \qty{160}{\degreeCelsius}.
        
        Samples are exposed with a \qty{100}{\kilo \electronvolt} electron-beam lithography (EBL) tool using a \qty{1100}{\micro \coulomb \per \cm \squared} primary dose and a \qty{170}{\micro \coulomb \per \cm \squared} undercut dose.

        After exposure, the samples are individually developed for \qty{60}{\s} in amyl acetate (AR 600-546) at \qty{0}{\degreeCelsius}, \qty{30}{\s} in isopropanol, \qty{600}{\s} in a $3:1$ mixture of isopropanol to methyl isobutyl ketone (MIBK) including agitation for the first and last \qty{60}{\s}, and a final \qty{30}{\s} rinse in isopropanol.
        If not otherwise specified, solvents are used at room temperature.
        Prior to aluminum deposition, we use oxygen plasma ashing (\qty{40}{\W} RF power, \qty{30}{\cm \cubed \per \minute} \ce{O2} flow, \qty{2}{\Pa} chamber pressure) in a reactive-ion-etching tool to clean the samples.
        
        We install the samples into an ultra-high-vacuum electron-beam evaporation system, and evaporate aluminum from three angles.
        In the first step, we deposit \qty{30}{\nm} of aluminum at a rate of \qty{0.5}{\nm \per \s} parallel with one junction lead at an incidence angle of \qty{45}{\degree} to the normal and then statically oxidize for \qty{180}{\s} in a \qty{15}{\percent} oxygen-argon mixture at \qty{400}{\Pa} in a separate chamber.
        To form the top contact, we rotate the sample stage \qty{90}{\degree} so that we are parallel with the orthogonal junction lead and then deposit \qty{40}{\nm} of aluminum twice, tilting the stage \qty{90}{\degree} in between the evaporations such that we deposit from both directions collinear with the second lead.
        We then oxidize the completed structure for \qty{300}{\s} in the same gas mixture at \qty{2}{\kPa}.

        After junction deposition, we lift off the unwanted aluminum over \qty{2}{\hour} in \qty{50}{\degreeCelsius} acetone.
        We clean the samples for \qty{30}{\min} in \qty{80}{\degreeCelsius} \emph{N}-methyl-2-pyrrolidone (NMP) followed by sonication for \qty{5}{\min} in \qty{50}{\degreeCelsius} NMP, acetone, and then isopropanol to remove residual resist.
        
        For the bandages, we again use a bilayer resist stack, starting with MMA 8.5 MAA 12 EL as before, but using a \qty{4}{\percent} polymethyl methacrylate (PMMA) 950k solution in EL as the top layer, spinning at \qty{4000}{\rpm} (\qty{4}{\s} ramp, \qty{60}{\s} hold) and baking for \qty{480}{\s} at \qty{160}{\degreeCelsius}.
        We also add a charge dissipating layer (Espacer 300Z), spinning at \qty{2000}{\rpm} (\qty{6}{\s} ramp, \qty{60}{\s} hold), and baking for \qty{90}{\s} at \qty{80}{\degreeCelsius}.

        We expose the samples using a \qty{30}{\keV} EBL system using a \qty{400}{\micro \coulomb \per \cm \squared} primary dose and a \qty{60}{\micro \coulomb \per \cm \squared} undercut dose.
        After exposure, we rinse the samples for \qty{30}{\s} in distilled water to remove the espacer and then develop for \qty{50}{\s} in a $3:1$ isopropanol MIBK mixture followed by \qty{30}{\s} in isopropanol.

        We place the samples in a high-vacuum electron-beam evaporation system, argon ion mill for \qty{90}{\s} (\qty{10}{\mA} current, \qty{400}{\V} beam voltage) at a \qty{30}{\degree} angle of incidence from normal while rotating the sample stage.
        To ensure that we fully cover the niobium film edge with the aluminum, we deposit \qty{300}{\nm} of aluminum at a \qty{10}{\degree} angle from normal with rotation.
        Before opening the chamber, we oxidize for \qty{600}{\s} in \ce{O2} at \qty{1.3}{\kPa} to grow a controlled surface oxide.
        We repeat the same lift-off and cleaning steps as for the junctions.

        \subsection{Indium deposition}
            \label{sec:appendix:fabrication:indium}

        All aspects of the indium fabrication are as discussed in Ref.~\cite{norris2024improved} except for the deposition itself which we  describe here.
        We load the bottom wafer and four-top-chip dies into a dedicated indium thermal evaporator.
        We ion-mill with a beam voltage of \qty{600}{\V}, a beam current of \qty{120}{\mA}, and an argon flow of \qty{20}{\cm \cubed \per \minute} for \qty{120}{\s} to remove the native niobium oxide.
        We evaporate \qty{10}{\um} of indium on all chips at a rate of \qty{1}{\nm \per \second} with the substrate cooling set to \qty{-5}{\degreeCelsius}.
        
        After dicing and protection layer removal, the samples are flip-chip bonded using a force of \qty{21.6}{\N} and then packaged as described in Ref.~\cite{norris2024improved}.

    \section{Measurement apparatus}
        \label{sec:appendix:measurementapparatus}

        \begin{figure}
            \centering
            \includegraphics[width=85mm]{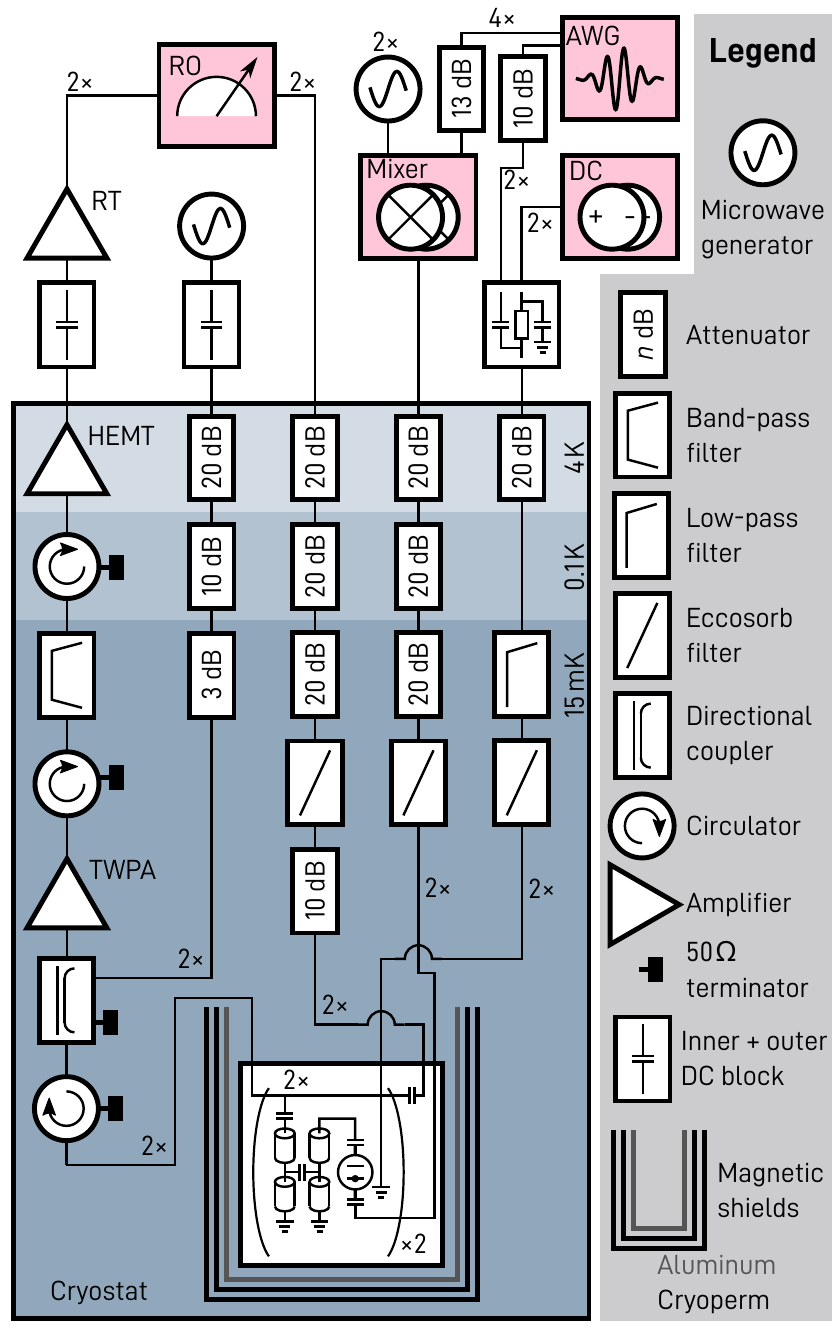}
            \caption{
                Wiring diagram of the measurement apparatus used for device A.
            }
            \label{fig:appendix:wiring-diagram}
    	\end{figure}

        We install each sample inside magnetic shielding at the base temperature stage of a commercial dilution refrigerator and connect the sample to room-temperature control electronics as indicated in Fig.~\ref{fig:appendix:wiring-diagram}.
        This diagram depicts the apparatus used to measure device A; for device B, an alternative apparatus with minor differences was used (discussed below).
        The successively colder cryostat stages feature attenuators to condition the incoming signals; these particular values were chosen based on Ref.~\cite{krinner2019engineering}.
        
        We control the qubit frequency by changing the magnetic flux threading the SQUID loop using a static current and base-band pulses.
        The qubits are tuned to idling frequencies using a static current created by an isolated voltage source.
        To change the qubit frequencies on nanosecond timescales, we apply current pulses created using an arbitrary waveform generator (AWG).
        We pre-distort the AWG waveforms to account for the transfer function of the cryostat wiring \cite{hellings2025calibrating}.
        We combine the static and base-band currents using an $RC$ bias-tee.

        We generate qubit XY drive pulses by creating $I$ and $Q$ quadrature pulse waveforms at an intermediate frequency of up to several hundred \si{\MHz} using an AWG and convert them up to the qubit drive frequency using an $IQ$ mixer and a microwave generator (MWG).

        Readout probe signals are generated by an integrated readout device combining a superheterodyne signal generator with a digitizer \cite{herrmann2022frequency}, and then applied to the readout feedline.
        After passing through the feedline, the readout signal is amplified by a traveling-wave parametric amplifier (TWPA) \cite{macklin2015near}, high-electron mobility amplifier (HEMT), and room-temperature (RT) low-noise amplifiers before being digitized by the acquisition device.
        
        For device B, the measurement apparatus is similar, but with a few minor differences.
        As the setup is intended for isolated qubit measurements, no AWGs are connected to the bias-tees on the flux lines.
        Within the cryostat, the readout input lines only feature \qty{20}{\dB} of attenuation at the base temperature stage; the eccosorb filters are placed prior to the \qty{20}{\dB} attenuators at the base stage; the TWPA pump line features an eccosorb filter at base instead of the \qty{3}{\dB} attenuator.
        The available evidence does not point to a clear advantage or disadvantage from these differences in signal conditioning.
        Due to a different hardware configuration in the mixing module, no attenuation is used between the charge drive AWGs and the mixer; \qty{20}{\dB} of attenuation and a splitter is instead placed between the output of the mixing module and the cryostat wiring.
    
    \section{Hamiltonian parameters}
        \label{sec:appendix:hamiltonian-parameters}

        \begin{table}[b]
            \centering
            \caption{
                Device A measured and extracted parameters of the Hamiltonian (Eq.~\ref{eq:appendix:hamiltonian-parameters:hamiltonian}).
                $\omega_\mathrm{q,ge}^\mathrm{d}(\phi)$, dressed qubit $\ket{\mathrm{g}}$-$\ket{\mathrm{e}}$ transition frequency at reduced flux $\phi = 2\pi \Phi / \Phi_0$ (with $\Phi$ the flux threading the SQUID loop and $\Phi_0$ the flux quantum); $\alpha_\mathrm{q}(\phi) = \omega_\mathrm{q,ef}^\mathrm{d}(\phi) - \omega_\mathrm{q,ge}^\mathrm{d}(\phi)$, qubit anharmonicity; $\omega_\mathrm{r}^\mathrm{b}$, bare resonator frequency.
                Extracted parameters: $E_\mathrm{j,max}$, Josephson energy at $\phi = 0$; $E_\mathrm{c}$, charging energy; $g_\mathrm{qr}$, qubit-resonator coupling rate.
            }
            \label{tab:appendix:hamiltonian-parameters:device-a}
            \begin{tabular}{l S S S S l} \toprule
                 & \multicolumn{4}{c}{{Qubit}} & \\ \cmidrule{2-5}
                {Parameter} & {1} & {2} & {3} & {4} & {Unit} \\ \midrule
                $\omega_\mathrm{q,ge}^\mathrm{d}(0.0)/2\pi$ &  5.415 &  5.662 &  5.584 &  5.008 & [\si{\GHz}] \\
                $\omega_\mathrm{q,ge}^\mathrm{d}(0.5)/2\pi$ &  4.421 &  4.736 &  4.411 &  3.585 & [\si{\GHz}] \\
                $\alpha_\mathrm{q}^\mathrm{d}(0.0)/2\pi$    & -0.159 & -0.158 & -0.165 &        & [\si{\GHz}] \\
                $\omega_\mathrm{r}^\mathrm{b}(0.0)/2\pi$    &  6.636 &  7.022 &  6.826 &  7.226 & [\si{\GHz}] \\ \midrule
                $E_\mathrm{j,max}/h$                        & 25.44  & 27.79  & 25.94  &        & [\si{\GHz}] \\
                $E_\mathrm{c}/h$                            &  0.154 &  0.154 &  0.161 &        & [\si{\GHz}] \\
                $g_\mathrm{qr}/2\pi$                        &  0.108 &  0.117 &  0.115 &        & [\si{\GHz}] \\
                \bottomrule
            \end{tabular}
        \end{table}
        
        We measure the extremal dressed qubit $\ket{\mathrm{g}}$-$\ket{\mathrm{e}}$ and $\ket{\mathrm{e}}$-$\ket{\mathrm{f}}$ transition frequencies from Ramsey measurements as a function of the flux bias applied to the qubit.
        We further measure the dressed readout-resonator and Purcell-filter coupled frequencies from spectroscopy measurements as a function of flux bias.
        Using high-power spectroscopy measurements \cite{blais2021circuit} with the qubits tuned to their minimum frequencies, we extract the bare readout-resonator frequencies by fitting the spectra as described in Appendix~\ref{sec:appendix:readout-parameters}.
        These values are presented in the upper half of Table~\ref{tab:appendix:hamiltonian-parameters:device-a}.
        We could not reliably find the $\ket{\mathrm{e}}$-$\ket{\mathrm{f}}$ transition of qubit 4 on device A and have thus excluded any values which depend on this from the table.

        With these measured and fit parameters, we then extract the device parameters numerically.
        We treat the coupled transmon-qubit--readout-resonator--Purcell-filter system as follows:
        \begin{equation}
            \label{eq:appendix:hamiltonian-parameters:hamiltonian}
            \begin{split}
            \hat{H} & = 4E_\mathrm{c} \hat{n}^2 - E_\mathrm{J} \cos{\hat{\phi}} + \hbar \omega_\mathrm{r} \hat{a}^\dagger \hat{a} + \hbar \omega_\mathrm{p} \hat{b}^\dagger \hat{b} \\
            & + i \hbar g_\mathrm{qr} \hat{n} \left(\hat{a}^\dagger - \hat{a}\right) - \hbar J_\mathrm{rp}\left(\hat{a} - \hat{a}^\dagger\right) \left(\hat{b} - \hat{b}^\dagger\right)
            \end{split}
        \end{equation}
        where $E_\mathrm{c}$ is the transmon charging energy, $E_\mathrm{J}$ is the transmon Josephson energy, $\hat{n}$ is the charge operator for the transmon, $\cos{\hat{\phi}}$ is the phase operator of the transmon, $\omega_\mathrm{r}$ is the readout-resonator frequency, $\hat{a}^\dagger$ ($\hat{a})$ is the resonator creation (annihilation) operator, $\omega_\mathrm{p}$ is the Purcell-filter frequency, $\hat{b}^\dagger$ ($\hat{b})$ is the filter creation (annihilation) operator, $g_\mathrm{qr}$ is the Jaynes--Cummings coupling between the qubit and resonator, and $J_\mathrm{rp}$ is the coupling between the resonator and filter.
        By numerically computing the eigenvalues of Eq.~\ref{eq:appendix:hamiltonian-parameters:hamiltonian} and iterating until the differences to the measured values are minimized, we find the parameters given in the lower half of Table~\ref{tab:appendix:hamiltonian-parameters:device-a}.

        We provide the relaxation times of qubits 3 and 4 on device A and qubits 1 and 2 on device B in Table~\ref{tab:appendix:relaxation-times}.
        Note that these are individual measurements, not averaged over time as in Sec.~\ref{sec:main:qubit-frequency-coherence}.
        The $T_2$ times of the qubits on device B are much lower than typical since they were measured away from the frequency extrema of the qubits.
        Device B was also exposed to air significantly longer than device A which may have reduced its $T_1$ times \cite{verjauw2021investigation}.
        
        \begin{table}
            \centering
            \caption{
                Device A and B relaxation times.
                The symbols are: $\omega_\mathrm{meas}$, $\ket{\mathrm{g}}$-$\ket{\mathrm{e}}$ transition frequency at which the relaxation time measurements were conducted; $T_{1}$, $\ket{\mathrm{e}}$-state longitudinal relaxation time; $T_{2}^{*}$, $\ket{\mathrm{e}}$-state transverse relaxation time under inhomogeneous broadening; $T_{2}^\mathrm{echo}$, $\ket{\mathrm{e}}$-state transverse relaxation time with a Hahn echo pulse.
                Note that these are from individual measurements rather than averaged over time.
            }
            \label{tab:appendix:relaxation-times}
            \begin{tabular}{l S S S S l} \toprule
                 & \multicolumn{2}{c}{{Dev. A Qubit}} & \multicolumn{2}{c}{Dev. B Qubit} & \\ \cmidrule{2-3} \cmidrule{4-5}
                {Parameter} & {3} & {4} & {1} & {2} & {Unit} \\ \midrule
                $\omega_\mathrm{meas}/2\pi$ &  5.584 &  3.585 &  4.454 &  5.894 & [\si{\GHz}] \\
                $T_1$                       & 37   & 53   & 57   & 41   & [\si{\us}]  \\
                $T_{2}^{*}$                 & 47   & 18   &  1.8   &  6.1   & [\si{\us}]  \\
                $T_{2}^\mathrm{echo}$       & 62   & 71   &  4.8   & 22   & [\si{\us}]  \\
                \bottomrule
            \end{tabular}
        \end{table}

    \section{Readout parameters}
        \label{sec:appendix:readout-parameters}

        To determine the readout parameters of each qubit, we measure the complex feedline transmission as a function of qubit initial state, as shown for device B in Fig.~\ref{fig:main:readout-performance}a,b.
        We then fit a model of the coupled readout-resonator--Purcell-filter system \cite{swiadek2024enhancing} to the complex feedline transmission spectra to extract the decoupled readout-resonator and Purcell-filter frequencies and other parameters shown in Table~\ref{tab:appendix:readout-parameters} for devices A and B.
        In the table: $\omega_\mathrm{p}$ refers to the frequency of the Purcell-filter, $\omega_\mathrm{r,g}$ is the frequency of the dressed readout-resonator with the qubit in the ground state, $\Delta_\mathrm{rp,g} = \omega_\mathrm{r,g} - \omega_\mathrm{p}$ is the detuning between the two modes, $J_\mathrm{rp}$ is the coupling rate between the resonator and filter, $\kappa_\mathrm{p}$ is the decay rate from the Purcell-filter into the feedline, $\kappa_\mathrm{r}$ is the effective decay rate from the readout-resonator into the feedline \cite{heinsoo2018rapid}, and $\chi$ is the dispersive shift \cite{blais2021circuit}.
        For device A, the parameters in Table~\ref{tab:appendix:readout-parameters} are with the qubits tuned to their operational frequencies (Fig.~\ref{fig:main:qubit-frequencies}a), while,
        for device B, the parameters in Table~\ref{tab:appendix:readout-parameters} are with qubit 1 (2) tuned to \qty{5.00}{\GHz} (\qty{5.89}{\GHz}).

        \begin{table}
            \centering
            \caption{
                Device A and B readout parameters.
                See text for a description of the variables.
            }
            \label{tab:appendix:readout-parameters}
            \begin{tabular}{l S S S S l} \toprule
                & \multicolumn{2}{c}{{Device A Qubit}} & \multicolumn{2}{c}{{Device B Qubit}} & \\ \cmidrule{2-3} \cmidrule{4-5}
                {Parameter} & {1} & {2} & {1} & {2} & {Unit} \\ \midrule
                $\omega_\mathrm{p}/2\pi$   &   6.699  &   7.107  &   6.707  &   7.116  & [\si{\GHz}] \\
                $\omega_\mathrm{r,g}/2\pi$ &   6.646  &   7.044  &   6.728  &   7.097  & [\si{\GHz}] \\ \midrule
                $\Delta_\mathrm{rp}/2\pi$  & -53.0 & -62.0 &  21.0 & -19.0 & [\si{\MHz}] \\
                $J_\mathrm{rp}/2\pi$       &  24.6 &  17.7 &  16.0 &  19.3 & [\si{\MHz}] \\
                $\kappa_\mathrm{p}/2\pi$   &  24.0 &  28.4 &  37.7 &  43.4 & [\si{\MHz}] \\
                $\kappa_\mathrm{r}/2\pi$   &   3.2 &   1.8 &   7.5 &  11.2 & [\si{\MHz}] \\
                $2\chi/2\pi$               &  -2.2 &  -5.6 &  -4.4 &  -6.2 & [\si{\MHz}] \\
                \bottomrule
            \end{tabular}
        \end{table}

        For device A, we measure qubit 1 (2) for \qty{320}{\ns} (\qty{240}{\ns}), recovering mean three-state assignment errors of \num{1.4e-2} (\num{1.8e-2}).
        In these measurements, we use a two-step readout pulse with greater initial amplitude (as discussed in the main text), use bichromatic \cite{chen2023transmon} readout for qubit 2 to overcome the small $\kappa_\mathrm{r}$ relative to $\chi$, and flux-pulse qubit 2 to a $\ket{\mathrm{g}}$-$\ket{\mathrm{e}}$ transition frequency of \qty{5.33}{\GHz} during the measurement to avoid a defect mode.
        For device B, the readout duration and resulting fidelities are discussed in Sec.~\ref{sec:main:readout} of the main text.
        For qubit 2, we reduce the readout duration by using a two-step pulse waveform to populate the readout resonator faster \cite{jeffrey2014fast}.
        
        We attribute the greater readout error of device A compared to device B to the large detuning between the readout modes, resulting in a small $\kappa_\mathrm{r}$ which limits readout speed.
        In addition, we find a quantum measurement efficiency \cite{bultink2018general} of \num{0.17} for qubit 2 of device A, compared to \num{0.29} for device B, indicating deficiencies in the cryogenic wiring or amplifiers used for the device A measurements.
        We further note for completeness that, for device A, a similar pre-selection as described in the main text discards \qty{3.8}{\percent} (\qty{0.7}{\percent}) of the data found not in $\ket{\mathrm{g}}$ for qubit 1 (2), which we attribute to the use of a sample mount with limited infrared-radiation screening for device A and a nearly light-tight one for device B \cite{gordon2022environmental,serniak2018hot}.

    \end{appendices}

    \section{Data availability}

        The data produced in this work are available from the corresponding author upon reasonable request.

	\section{Acknowledgments}

		We thank the staff of the ETH Zurich FIRST, the Binnig and Rohrer Nanotechnology Center, and the PSI PICO cleanrooms for their assistance in maintaining the facilities used to prepare the devices discussed in this work.
        We also thank MIT Lincoln Laboratory for supplying the traveling wave parametric amplifiers used in this work.

		The authors acknowledge financial support by the Swiss State Secretariat for Education, Research and Innovation under contract number UeM019-11, by the Intelligence Advanced Research Projects Activity (IARPA) and the Army Research Office, under the Entangled Logical Qubits program and Cooperative Agreement Number W911NF-23-2-0212, by the Baugarten Foundation, by the ETH Zurich Foundation, and by ETH Zurich.
        The views and conclusions contained in this document are those of the authors and should not be interpreted as representing the official policies, either expressed or implied, of IARPA, the Army Research Office, or the U.S.\ Government.
        The U.S.\ Government is authorized to reproduce and distribute reprints for Government purposes notwithstanding any copyright notation herein.

    \section*{Author Contributions}

		G.J.N., K.D., and J.-C.B.\ planned the experiments.
		K.D.\ designed and K.D., D.C.Z., A.R., A.F., and M.B.P.\ fabricated the devices.
        F.S., C.S., and C.H. assembled one of the experimental setups used in this work.
		G.J.N.\ and K.D.\ performed all measurements and analyzed all data.
		G.J.N.\ and K.D.\ wrote the manuscript with input from all authors.
		J.-C.B.\ and A.W.\ supervised the work.

    \section*{Additional information}

        \subsection*{Competing interests}

            The authors declare no competing interests.

        \subsection*{List of abbreviations}

        \begin{itemize}[label = {}]
            \item 3D: Three-dimensional
            \item CZ: Controlled-Z
            \item SQUID: Superconducting Quantum Interference Device
            \item CDF: Cumulative Distribution Function
            \item TWPA: Travelling-wave Parametric Amplifier
            \item RB: Randomized Benchmarking
            \item IRB: Interleaved Randomized Benchmarking
            \item EPC: Error per Clifford
            \item EPG: Error per Gate
        \end{itemize}
        \begin{itemize}[label = {}]
            \item MMA: Methyl Methacrylate
            \item MAA: Methacrylic Acid
            \item EL: Ethyl Lactate
            \item EBL: Electron-beam Lithography
            \item MIBK: Methyl Isobutyl Ketone
            \item NMP: N-Methyl-2-Pyrrolidone
            \item PMMA: Polymethyl Methacrylate
            \item AWG: Arbitrary Waveform Generator
            \item MWG: Microwave Generator
            \item HEMT: High Electron-Mobility Amplifier
            \item RT: Room Temperature
        \end{itemize}
        
    \bibliography{references}

\end{document}